# CALYPSO: a method for crystal structure prediction

Yanchao Wang, Jian Lv, Li Zhu and Yanming Ma*

*State Key Laboratory of Superhard Materials, Jilin University, Changchun 130012, China*

We have developed a software package CALYPSO *(Crystal structure AnaLYsis by Particle Swarm Optimization)* to predict the energetically stable/metastable crystal structures of materials at given chemical compositions and external conditions (e.g., pressure). The CALYPSO method is based on several major techniques (e.g. particle-swarm optimization algorithm, symmetry constraints on structural generation, bond characterization matrix on elimination of similar structures, partial random structures per generation on enhancing structural diversity, and penalty function, etc) for global structural minimization from scratch. All of these techniques have been demonstrated to be critical to the prediction of global stable structure. We have implemented these techniques into the CALYPSO code. Testing of the code on many known and unknown systems shows high efficiency and high successful rate of this CALYPSO method [Wang *et al.*, Phys. Rev. B 82 (2010) 094116][1]. In this paper, we focus on descriptions of the implementation of CALYPSO code and why it works.

## 1. Introduction

Understanding the behaviors of materials at the atomic scale is fundamental to modern science and technology. As many properties and phenomena are ultimately controlled by the crystal structures, the prediction of crystal structure is an important task in chemistry and condensed matter physics. However, the structural prediction with the only known information of chemical compositions is extremely difficult as it basically involves in classifying a huge number of energy minima on the lattice energy surface. Owing to the significant progress in both computational power and basic materials theory, it is now possible to predict the crystal structure at zero Kelvin using the quantum mechanical methods. One way to predict structure is by extracting known structures from databases of structures previously found in similar materials[2]. However, this method has a limited success rate and is incapable of generating new crystal structure types. Recently, the more advanced methods including simulated annealing[3, 4], minima hopping[5], basin hopping[6], metadynamics[7], genetic algorithm[8-15], and random sampling method[16]have been developed and applied, which allow a systematic search for the ground state structures based on the chemical composition and the external conditions. The simulated annealing, basin hopping, minima hopping and metadynamics focus on overcoming the energy barriers and are successful in many researches[3-7], particularly, when the starting structure is close to the global minimum. The genetic algorithm starts to use a self-improving method and is thus able to correctly predict many structures [17-20]. The random sampling method, as a simple and efficient method, is also successful in many applications [21-24].

The particle swarm optimization(PSO), first proposed by Kennedy and Eberhart [25, 26], is a population-based optimization method. As a stochastic global optimization method, PSO is inspired by the choreography of a bird flock and can be seen as a distributed behavior algorithm that performs multidimensional search. According to PSO, the behavior of each individual is affected by either the best local or the best global individual to help it fly through a hyperspace. Moreover, an individual can learn from its past experiences to adjust its flying speed and direction.

Therefore, all the individuals in the swarm can quickly converge to the global position. PSO algorithm is a highly efficient global optimization method which has been applied successfully into many optimization problems such as network training [27, 28] and transactions on power systems[29]. However, the application of PSO to the structural prediction of condensed matters remains a major challenge. Due to the existence of a large number of energy minima on the lattice energy surface, rapid swarm convergence, as one of the main advantages of PSO, can also be problematic. If an early solution is sub-optimal, the swarm can easily stagnate around it without any pressure to continue further exploration, i.e., the premature. We recently have developed a CALYPSO method/code[1] on crystal structure prediction by implementation of PSO algorithm and many other important techniques, including symmetry constraints on structural generation, bond characterization matrix on elimination of similar structures, partial random structures per generation on enhancing structural diversity, and penalty function, etc. We found that these later techniques are critical to avoid the premature of PSO algorithm and to significantly accelerate the structure convergence.

The description of CALYPSO method and its first applications to the prediction of crystal structures can be found in Ref.[1]. This paper is organized as follows. The detailed descriptions of implementation of CALYPSO code and the principles on illustrating why the method works are presented in Section 2. In Section 3, various parameters in CALYPSO code are optimized for $TiO_2$ as a benchmark. The input and output files are provided in Sections 4. A short overview of the applications obtained from our method can be found in Section 5, followed by the conclusion in Section 6.

## 2. Implementation and discussions

As depicted in the pseudo-code of Algorithm1, the CALYPSO method comprises mainly four steps: (i) generation of random structures with the constraint of symmetry; (ii) local structural optimization; (iii) post-processing for the identification of unique local minima by bond characterization matrix; (iv) generation of new structures by PSO for iterations.

### 2.1 Symmetry constraints on structural generation

There are two types of variables to define a crystal structure: lattice parameters (three angles and the lengths of the three lattice vectors) and atomic coordinates (three coordinates coded as a fraction of the lattice vector for each atom). The first step of CALYPSO method is to generate random structures constrained within 230 space groups. Once a particular space group is selected, the lattice parameters are generated within the chosen symmetry according to the confined volume and the corresponding atomic coordinates are obtained by a combination of a set of symmetrically related coordinates (Wyckoff Positions) in accordance to the number of atoms in the simulation cell. For example, if the confined volume is 64 $Å^3$ and there are 12 atoms in the simulation cell for the group 223(Pm-3n), the lengths of three lattice vectors should be 4 Å and the lattice angles are fixed to 90°, while the atomic positions can be combined by different Wyckoff Positions (e.g., 6b + 6c, 6b + 6d, and 12f, etc). Moreover, a list including the symmetric information (space group) of all generated structures is built and used to compare with the newly generated structures. The appearance of identical symmetric structures is forbidden with a certain probability (80%). This makes the initial sampling covered different regions of the search space, which is crucial for the diversity of population. The generation of random structures ensures unbiased sampling of the energy landscape. The explicit application of symmetric constraints leads to significantly reduced search space and optimization variables, and thus fastens global structural convergence.

In order to examine the efficiency of symmetric constraints as implemented in CALYPSO code, the system of $TiO_2$with 16 $TiO_2$ units (48 atoms) per simulation cell was used as a test case. 3250 structures at ambient pressure were randomly generated and then structurally optimized using the GULP code[30] with a combination of Buckingham and Lennard-Jones potentials[11, 31]. Fig. 1 (a) and (b) show the energy distributions of these generated structures with and without symmetry constraints, respectively. It is found that the rutile structure, i.e., the global stable structure cannot be generated if without symmetry constraints. However, once the symmetry is implemented in the generation of random structures, 203 (~6.2% in total) rutile structures were successfully produced. In order to further compare the structural search efficiency of generation of random structures with or without the symmetry constraints, the binary Lennard-Jones crystal $A_2B$ (18 atoms per simulation cell) was used as another test case. 5000 structures were randomly generated and then structurally optimized using the GULP code[30] with Lennard-Jones potentials ($\sigma_{AA}$=$\varepsilon_{AA}$=1.0, $\sigma_{BB}$=0.88, $\varepsilon_{BB}$=0.5, $\sigma_{AB}$=0.932 and $\varepsilon_{AB}$=1.5)[32]. The energy distributions of the structures generated with and without symmetry constraints are shown in Fig. 1 (c) and (d), respectively. It is obvious that the energies of these structures generated with symmetry constraints distribute lower energy regions. We also have examined the structural search efficiency of CALYPSO runs with or without the symmetry constraints on structural generation as shown in Table 1. Obviously, the application of symmetry constraints technique can greatly improve the search efficiency, especially for larger systems. It is found that an averaged 11 generations are necessary to find the global stable structure if with the symmetry constraints on structural generation, however if without, 25.4 generations are needed. These tests clearly illustrate the importance of the symmetry constraints in the generation of random structures for structure prediction.

**2.2 Structural optimization**

CALYPSO code currently can use *ab initio* packages (e.g., VASP[33, 34], SIESTA[35] and CASTEP[36, 37]) and force-field program (e.g., GULP[30])to perform the structural optimization. Other external programs can also be interfaced on user's request. The use of locally structural optimization techniques (e.g., line minimization, steepest descents, conjugate gradient algorithm or Broyden-Fletcher-Goldfarb-Shanno algorithm) leads the lattice energy to the local minimum. Here, we use free energy (at T = 0 K, free energy reduces to enthalpy) as fitness function throughout the simulation. Note that local optimization increases the cost of each individual, but reduces effectively the noise of the energy landscape, enhances comparability between different structures, and provides locally optimal structures for further use. Thus, local optimization is crucial for the structure prediction.

**2.3 Elimination of similar structures by using the bond characterization**

**matrix**

Our goal is to eliminate the similar structures in the structure generations to enhance the search efficiency of CALYSPO. In our earlier implementation [1], we used geometrical structure parameter, which is solely based on the bond length, to identify structural similarity. Here, we have developed a more efficient technique named as bond characterization matrix, which is on the basis of all the bond information. In this method, we employ a set of modified bond-orientational order metrics($Q_l$) introduced by Steinhardt*et al*[38] to quantify the bond angles and an exponential function to quantify the bond length. When the distance between two atoms is less than the cutoff ($r_{cut}$), bond information, *e.g.* bond vector ($\vec{r}_{ij}$), bond angles ($\theta_{ij}$, $\phi_{ij}$) and bond-types ($\delta_{AB}$), are evaluated, where $\vec{r}_{ij}$ is a vector pointing from *i*thatom to *j*thatom, while $\theta_{ij}, \phi_{ij}$ are the related polar and azimuthal angles of $\vec{r}_{ij}$, respectively, and *A*(*B*)is the type of *i*th(*j*th) atom. In this work, bond characterization matrix is calculated according to the "bond-types", where each vector $\vec{r}_{ij}$ can be represented by spherical harmonics $Y_{lm}(\theta_{ij}, \phi_{ij})$. Subsequently, for each bond type$\delta_{AB}$, a weighted average is performed,

$$\overline{Q}_{lm}^{\delta_{AB}} = \frac{1}{N_{\delta_{AB}}} \sum_{i\in A, j\in B} e^{-\alpha(r_{ij} - b_{AB})} Y_{lm}\left(\theta_{ij}, \phi_{ij}\right) \tag{1}$$

where $N_{\delta_{AB}}$ is the number of bonds formed by type *A* and *B* atoms, $b_{AB}$ is the shortest length for each bond type and $\alpha$ is an adjusted parameter driving $e^{-\alpha(r_{cut} - b_{AB})} \rightarrow 0$.In order to avoid the dependence on the choice of reference frame, the average $\overline{Q}_{lm}^{\delta_{AB}}$ is used to calculate the rotationally invariant combinations,

$$Q_l^{\delta_{AB}} = \sqrt{\frac{4\pi}{2l+1} \sum_{m=-l}^{l} \left| \overline{Q}_{lm}^{\delta_{AB}} \right|^2} \tag{2}$$

Only even-l spherical harmonics, which are invariant with respect to the direction of the bonds, are used in Eq. (2), and each structure can be characterized by bond characterization matrix. The similarity between two structures is thus given by the Euclidean distance of their bond characterization matrix.

$$D_{uv} = \left[ \sum_{\delta_{AB}} \sum_{l} (Q_l^{\delta_{AB},u} - Q_l^{\delta_{AB},v})^2 \right]^{1/2} \quad (3)$$

where u and v are individual structures.

As an illustrative case, the histograms of $Q_l$ versus $l$ for graphite and diamond are shown in Fig. 2 (a) and (b), respectively. Significant differences for $Q_l$ between these two structures are evidenced, which illustrate the efficiency of the bond characterization matrix method to distinguish different structures. To further demonstrate the robust of the method, the Euclidean distances between graphite/diamond and its random distortions are calculated as shown in Fig. 2 (c)/(d).It is clearly seen that the calculated Euclidean distances monotonously increase with the magnitude of distortions. These tests highlight the capability of this bond characterization matrix method in the characterization of the structural similarities.

We have implemented this bond characterization matrix technique into CALYPSO code to eliminate similar structures. Table 2 shows the influence of this technique on the search efficiency of CALYPSO calculations for the system of $TiO_2$. It is clearly seen that much fewer optimization steps are needed to find the stable structure when this technique is included in the CALYPSO runs. This is understandable since the use of bond characterization matrix technique can effectively avoid the presence of very similar or identical structures and thus is able to accelerate the global structure convergence.

### 2.4 Generation of new structures by PSO

Within the PSO scheme, a structure (an individual) in the searching space is regarded as a particle. A set of individual structures is called a population. The lattice parameters (unit cell) of new structures are the same as the corresponding structures of the previous generation. While the atomic positions are updated using the evolutionary equation (4). Note that all the new structures produced by PSO (or randomly generated) are tested against constraint of minimal inter-atomic distances[10].

$$x_{i,j}^{t+1} = x_{i,j}^{t} + v_{i,j}^{t+1} \quad (4)$$

The initial $v_{i,j}$ was generated randomly. According to equation (5), the new velocity ($v_{i,j}^{t+1}$) of each individual *i* at the *j*th dimension (X Y Z), is calculated based on the velocity of previous generation ($v_{i,j}^{t}$), its previous location ($x_{i,j}^{t}$) before structural optimization, current location ($pbest_{i,j}^{t}$) after structural optimization, and the population global location ($gbest_{i,j}^{t}$) with the best fitness value for the entire population. It is obvious that the velocity of PSO is different from the physical velocity. The velocity of PSO is generated by the atomic coordinates and other dimensionless parameters, so it has the same unit with the atomic position. It is noted that the velocity plays an important role on determination of the speed and direction of structural movement.

$$v_{i,j}^{t+1} = \omega v_{i,j}^{t} + c_1 r_1 (pbest_{i,j}^{t} - x_{i,j}^{t}) + c_2 r_2 (gbest_{i,j}^{t} - x_{i,j}^{t}) \quad (5)$$

where $j \in \{X, Y, Z\}$, $\omega$ denotes the inertia weight, $c_1$ and $c_2$ are self-confidence factor and swarm confidence factor. High settings of $\omega$ as 0.9 facilitate global search, and lower settings facilitate rapid local search. In our methodology, $\omega$ is dynamically varied and decreases linearly from 0.9 to 0.4 during the iteration according to equation (6).

$$\omega = \omega_{max} - \frac{\omega_{max} - \omega_{min}}{iter_{max}} \times iter \quad (6)$$

Where $\omega_{max}$ and $\omega_{min}$ equals to 0.9 and 0.4, respectively. Accordingly, in our implementation, $c_1$ and $c_2$ are kept as constant 2. $r_1$ and $r_2$ are two separately generated random numbers in the range 0 to 1. As shown in equation (5), it is quite obvious that the movement of particles in the search space is dynamically influenced by their individual past experience ($pbest_{i,j}^{t}$, $v_{i,j}^{t}$) and successful experiences attained by the whole swarm ($gbest^{t}$). Thus the velocity makes the particles to move towards to global minimum and accelerates the convergence speed. The settings of other parameters will be presented in Section 3.

### 2.5 Penalty function

According to Bell-Evans-Polanyi principle[16, 39], the low energy basins in potential energy surfaces are expected to occur near other low energy basins. Thus, in order to improve the efficiency of the procedure, a certain number of high-energy structures are rejected, and the remaining low energy structures, which are on the most promising areas of the configuration space, are selected to produce the next generation by PSO. Fig. 3 (a) and (b) show the evolution of lattice energy distributions with and without the inclusion of penalty function during the simulation (shown here for $TiO_2$ with 48 atoms in the simulation cell). Obviously, most of structures are in low-energy region (<620.0eV) when the penalty function technique is included and it significantly accelerates the structural converges to the global minimum as demonstrated in the CALYPSO runs (Fig. 3).

### 2.6 Structural diversity

Structural diversity plays an important role in the prediction of crystal structures by using the population-based methods, such as the genetic algorithm and our developed CALYPSO method. During the structural evolution, if the systems lose the structural diversity, it is quite often that the systems stagnate, particularly for a large system. We here have designed a critical technique to enhance the structural diversity by including certain percentage of random structures in each generation, which has been implemented in CALYPSO code. Again, we use $TiO_2$ with 16 formula units per simulation cell as a test example. The history of CALYPSO runs with and without including the randomly generated structures is shown in Fig. 4. It is seen that the inclusion of a certain number of structures whose symmetries must be distinguished from any of previously generated ones is indeed crucial to converge the system to the global minimum. This all comes to the true fact that the inclusion of random structures allows the generation of diverse structures [Table 3]. Note that it might come up with the question on if the global stable structure is in fact generated by those random structures. We have performed a certain number of tests and found out that only a few stable structures are generated randomly, especially for smaller systems. For most of cases, the structural evolution of CALYPSO runs derives the global stable structures.

**2.7Convergence**

The CALYPSO simulation is stopped when the halting criterion is reached. In accordance with our experience, the stable crystal structure can usually be found at ~10 generations for systems ≤10 atoms per simulation cell. In practice, the halting criterion in CALYPSO is by default set to10 further generations if the simulation can not find other better structures.

## 3. Optimization of parameters

In order to provide reasonable default setting for various parameters in our CALYPSO code, a test was performed on $TiO_2$ system with 16 formula units per simulation cell by using the GULP code for the structural optimization and total energy calculations. Earlier study [26] has demonstrated that $c_1 = c_2 = 2$ and the linear decrease of ω from 0.9 to 0.4 during the iteration usually give the best overall performance for PSO simulations. Thus, we adopt these parameters and other parameters such as the population size ($N_{POP}$), the proportion of the structures generated by PSO($P_{PSO}$) and the max magnitudes of the velocity ($V_{max}$) are determined by using the benchmark of $TiO_2$. We repeat 5 successful CALYPSO calculations, i.e., the correct finding of rutile structure, to derive the proper parameters. The results and suggested parameter values can be found in Table 4.

## 4. Input and output files

4.1 input file

The main input file named as input.dat, contains all the necessary parameters for the simulation. There are several examples for the input.dat file in the Examples directory of CALYPSO package.

We here take SiC as an example:

SystemName = SiC

NumberOfSpecies = 2

NameOfAtoms = C Si

NumberOfAtoms = 1 1

NumberOfFormula = 2 2

AtomicNumber = 6 14

MaxStep = 50

Volume= 20.0

@DistanceOfIon

1.2 1.5

1.5 1.9

@End

PsoRatio = 0.6

Icode= 1

Kgrid = 0.12 0.08

Command = vasp

PopSize = 20

PickUp = F

PickStep = 0

Here follows a description of the variables defined in the input file (input.dat), including the data types and default values.

SystemName (string): A string of one or several words contains a descriptive name of the system (max. 40 characters).

Defualt value: CALYPSO

NumberOfSpecies (integer): Number of different atomic species.

Default value: No default.

NameOfAtoms (string): Element symbols of the different chemical species.

Default value: No default.

AtomicNumber (integer): Atomic Number of each chemical species.

Default value: No default.

NumberOfAtoms (integer): Number of atoms for each chemical species in one formula unit.

Default value: No default.

NumberOfFormula (integer): The desired range of formula units per simulation cell. The first and second numbers are the lower and upper limits per simulation cell in the formula units.

Default value: 1    4

Volume (real): The volume per formula unit. Unit is in $Å^3$. The volume can be estimated by the atomic volume of given elements. If it is set to zero, the program will automatically generate the estimated volume by the radius of ions.

Default value: 0

@DistanceOfIon and @End (real): Minimal distances between different chemical species. Unit is in angstrom. The determination of this parameter is in accordance with " NumberOfSpecies". For example, if the NumberOfSpecies=2, a 2×2 matrix is used to indicate the minimal distances between different chemical species.

@DistanceOfIon

d11 d12

d21 d22

@End

Default value: 0.7 Å

Icode(integer): It determines which local optimization package should be interfaced with in the simulation.

1: VASP

2: SIESTA

3: GULP

4: CASTEP

Default value: 1

PsoRatio (real): The proportion of the structures generated by PSO, and the other structures will be generated randomly.

Default value: 0.6

PopSize (integer): The population size. Normally, it will have a larger value for larger systems.

Default value: 30

Kgrid (real): The precision of the K-point sampling for local optimization (VASP

or SIESTA). The Brillouin zone sampling uses a grid of spacing 2π×Kgrid Å-1. The first value controls the precision of the first two local optimizations, and the second value with denser K-points controls the last optimization. The smaller value normally gives finer optimization results.

Default value: 0.12    0.06

Command (string): The command to perform local optimization on your computer.

Default value: submit.sh.

MaxStep (integer): The maximum number of PSO iterations. It should have a larger value for a larger system.

Default value: 50

PickUp(logical): If True, a previous calculation will be continued.

Default value: false

PickStep(integer): At which step will the previous calculation be picked up.

Default value: There is no default. If PickUp=True, you must supply this variable.

4.2 output files

The main outputs of CALYPSO are in the "results" folder:

CALYPSO.log: It includes the information of the structures (the space group, the volume, the number of atoms, *et al*.).

similar.dat: It includes the bond characterization matrixes of predicted structures.

pso_ini_*: It includes the information of the initial structures of the *-th iteration step.

pso_opt_*: It includes the enthalpy and structural information after local optimization of the *-th iteration.

pso_sor_*: The enthalpy sorted in ascending order of the *-th iteration step.

## 5. Applications.

We have earlier illustrated that the CALYPSO method can be used to predict various structures on elemental, binary and ternary compounds with various chemical bonding environments (e.g., metallic, ionic, and covalent bonding)[1, 40, 41]. Here, we discuss some other applications on the discovery of hitherto unknown structures. All the *ab initio* structure relaxations were performed using density functional theory within the projector augmented wave method, as implemented in the VASP code [33, 34]. An overview of systems with unknown structures for which we have performed calculations and discovered new structures can be found in Table 5.

Lithium (Li) is a "simple" metal at ambient pressure, but exhibits complex phase transitions under compression. Experimentally, it has been demonstrated that Li takes the phase transition sequence of bcc→ fcc→ hR1 → cI16, above which new phases are observed but remain unsolved[42]. We thus have extensively explored the high-pressure phases of Li through CALYPSO code. We successfully predicted all the experimental structures at certain pressure ranges by the CALYPSO method[1]. In particular, two new orthorhombic *Aba*2-40 (40 atoms/cell) and *Cmca*-56(56 atoms/cell) structures of Li [43] were predicted at 80 and 200 GPa. These two complex structures (Aba2-40 and Cmca-56) are successfully predicted only at the third and fourth generation with a population size $N_{pop}$ of 30, respectively. Note that *Aba*2-40 (oC40) structure has been later verified by an independent experiment[44].

Being a known best thermoelectric material and a topological insulator at ambient condition, bismuth telluride experiences phase transitions into several superconducting states under pressure. However, the high-pressure structures remain unsolved since 1972. We have recently predicted two low-pressure phases of bismuth telluride through CALYPSO calculations as seven-fold ($\beta$-$Bi_2Te_3$) and eight-fold ($\gamma$-$Bi_2Te_3$) monoclinic structures at 12 and 14 GPa, respectively[45]. These two structures were identified at the first and fifth generation with a population size of 30 and 40. These structures also have been subsequently verified by our experiment through Reitveld refinement [45]. Other compounds (Mg[46], $BC_3$[47]and $BC_7$[48]) with unknown structures also are discovered at high pressure by CALYPSO simulations [Table 5]. All the structures rapidly converge to the global minimum with less than 150 local optimizations. These results demonstrated that our method is a powerful and efficient tool on crystal structure determination.

The reason why our method is so successful can be traced to several powerful techniques. Firstly, PSO is a highly efficient global optimization algorithm, which has been applied successfully into many multi-objective optimization problems. Secondly, symmetry constraints on structural generation make the initial sampling covered different regions of the search space, which is crucial for the efficiency of global minimization. Thirdly, the elimination of similar/identical structures using bond characterization matrix technique and rejection of high-energy structures for each generation are able to accelerate the global structural convergence. Fourth, the inclusion of a certain number of structures whose symmetries are distinguished from previous ones can keep the population diversity and is critical to the prediction of global stable structures. Finally, the local optimization is effective reduce the noise of the landscape and may also be one of the key issues for our method success.

## 6. Conclusions

In this paper, we outline descriptions of implementation of CALYPSO code, which can be used to predict crystal structures of materials at given chemical compositions and external conditions. Our CALYPSO method has incorporated several major techniques (e.g. PSO algorithm, symmetry constraints on structural generation, bond characterization matrix on elimination of similar structures, partial random structures per generation on enhancing structural diversity, and penalty function, etc), which have been demonstrated to be crucial to the prediction of global stable structure. Suggested values for various parameters in CALYPSO have been presented by performing benchmark on $TiO_2$ system. The high success rate and high efficiency on the structural searches of CALYPSO methodology have demonstrated its reliability and promise as a major tool on crystal structure determination.

## Program availability

CALYPSO is available via http://nlshm-lab.jlu.edu.cn/~calypso.html. The software is free of charge for non-profit organizations, and delivered with the Fortran source code. The details of installation instructions, the user's manual in PDF format and examples are included in the package.

## Acknowledgement

The authors acknowledge the funding supports from the National Natural

Science Foundation of China under grant Nos. 11025418 and 91022029.

*Correspondence and requests for materials should be addressed to Y.M. (mym@jlu.edu.cn).

## Table and Figure captions

**TABLE 1** The structural search efficiency of CALYPSO calculations with or without the symmetric constraints on structural generation for the system of $TiO_2$.We have performed ten different CALYPSO runs and the total generations for these ten runs needed to find the global stable rutile structure are listed. As an illustration, we choose here the population size as 20. Notably, we generally use larger population sizes for larger systems; there much less generations are needed to find the stable structure. Other typical CALYPSO run parameters of $V_{max}$ and the percentage of PSO generated structures are chosen as 0.1 and 0.6, respectively.

**TABLE 2** The structural search efficiency of CALYPSO calculations with or without the elimination of similar structures for the system of $TiO_2$.We have performed ten different CALYPSO runs and the total generations for these ten runs needed to find the global stable rutile structure are listed. As an illustration, we choose here the population size as 20. Notably, we generally use larger population sizes for larger systems; there much less generations are needed to find the stable structure. Other typical CALYPSO run parameters of $V_{max}$ and the percentage of PSO generated structures are chosen as 0.1 and 0.6, respectively.

**TABLE 3** The structural search efficiency of CALYPSO calculations with or without partial random structures per generation for the system of $TiO_2$.We have performed ten different CALYPSO runs and the total generations for these ten runs needed to find the global stable rutile structure are listed. As an illustration, we choose here the population size as 20. Notably, we generally use larger population sizes for larger systems; there much less generations are needed to find the stable structure. Other typical CALYPSO run parameter of $V_{max}$ is chosen as 0.1.

**TABLE 4** The test of variable parameters in CALYPSO.

**TABLE 5** Systems with unknown structures, for which we have done calculations

and revealed new structures.

**Algorithm 1.** The pseudo-code of the implementation of CALYPSO.

**FIG. 1. (color online)** The energy distributions of randomly generated structures containing 16 $TiO_2$ units in the simulation cell and 6 units of binary Lennard-Jones crystal $A_2B$ in the simulation cell after local optimization. (a) and (b) indicate the energy distribution of $TiO_2$ structures generated with and without the symmetric constraints, respectively. (c) and (d) indicate the energy distribution of $A_2B$ structures generated with and without the symmetric constraints, respectively.

**FIG. 2. (color online)**(a) and (b) $Q_l$ histograms for graphite and diamond structures, respectively. (c) and (d) distance against distortion for graphite and diamond structures, respectively. The unit of distortion magnitude is in bond length.

**FIG. 3. (color online)**(a) and (b) represent the evolution of lattice energy distributions during structural iterations with and without the inclusion of penalty function, respectively.

**FIG. 4. (color online)** The history of CALYPSO search performed on $TiO_2$ with 48 atoms per cell. The red line represents the CALYPSO runs on that a certain number of the low energy structures (0.6 of total) are selected to produce the next generation by PSO, while the rest of structures are generated randomly. The green line represents that all the structures are used to generate the next generation by PSO. Note that the stable structure is produced by PSO in these calculations.

# Table 1

| | Number of atoms in the system | | | |
|---|---|---|---|---|
| | 12 | 24 | 36 | 48 |
| | Generations | Generations | Generations | Generations |
| Symmetry constraints | 12 | 15 | 85 | 110 |
| No symmetry constraints | 12 | 25 | 138 | 254 |

# Table 2

| | Number of atoms in the system | | | |
|---|---|---|---|---|
| | 12 | 24 | 36 | 48 |
| | Generations | Generations | Generations | Generations |
| To eliminate similar structures | 11 | 14 | 21 | 78 |
| To preserve similar structures | 10 | 17 | 47 | 118 |

# Table 3

| | Number of atoms for $TiO_2$ | | | |
|---|---|---|---|---|
| | 12 | 24 | 36 | 48 |
| $P_{PSO}$ | Generations | Generations | Generations | Generations |
| 0.6 | 12 | 15 | 85 | 110 |
| 1.0 | 11 | 23 | 124 | 229/9[a] |

[a]It fails to find the global stable structure in 100 generations one time out of ten.

# Table 4

| | Test results | | | | | Suggested values |
|---|---|---|---|---|---|---|
| $P_{PSO}$ | 0.5 | 0.6 | 0.7 | 0.8 | 0.9 | 0.7-0.8 |
| Generations | 61/5 | 81/5 | 37/5 | 39/5 | 86/5 | |
| $V_{max}$ | 0.05 | 0.1 | 0.2 | 0.3 | 0.4 | 0.1-0.2 |
| Generations | 32/5 | 31/5 | 27/5 | 37/5 | 31/5 | |
| $N_{POP}$ | 10 | 20 | 30 | 40 | | 30 |
| Structures | 1480/5 | 2000/5 | 840/5 | 1600/5 | | |

# Table 5

| Systems | Pressure (GPa) | Structures | Generations | $N_{pop}$ |
|---|---|---|---|---|
| Li | 80 | *Aba*2-40[a] | *3* | *30* |
| | 200 | *Cmca*-56[a] | *4* | *30* |
| Mg | 500 | fcc[b] | 4 | 20 |
| | 800 | sh[b] | 5 | 30 |
| $Bi_2Te_3$ | 12 | β-$Bi_2Te_3$[c] | 1 | 40 |
| | 14 | γ-$Bi_2Te_3$[c] | 5 | 30 |
| | 20 | C2/m(bcc-like)[c] | 2 | 40 |
| $BC_3$ | 0 | *Pmma*[d] | *4* | *30* |
| $BC_7$ | 0 | *P*-4*m*2[e] | *6* | *20* |

[a]Ref. [43][b]Ref. [46][c]Ref.[45][d]Ref. [47][e]Ref.[48]

Number of particles, N; swarm, S; volume, V; Percentage of PSO generated structures, $P_{PSO}$.

Initialization of S (Generation of random structures with constraint of symmetry)

Evaluation of S (Local optimization) and definition of the *pbest* and *gbest*

List of the bond characterization matrixes (BCM)

```
While not done do
    S_PSO=S*P_PSO and S_random=S*(1-P_PSO)
    While i<=S_PSO do
        S(i)(Generation of new structures by PSO)
        If S(i) ∉BCM then
            i=i+1
            To update the list of BCM
        End if
    End while
    While i <=S_PSO+S_random
        S(i) Generation of random structures with constraints of symmetry
        If S(i) ∉BCM then
            i=i+1
            To update the list of BCM
        End if
    End while
    To Evaluate S (local optimization) and update the gbest
    To update the list of BCM
End while
```

# Algorithm 1

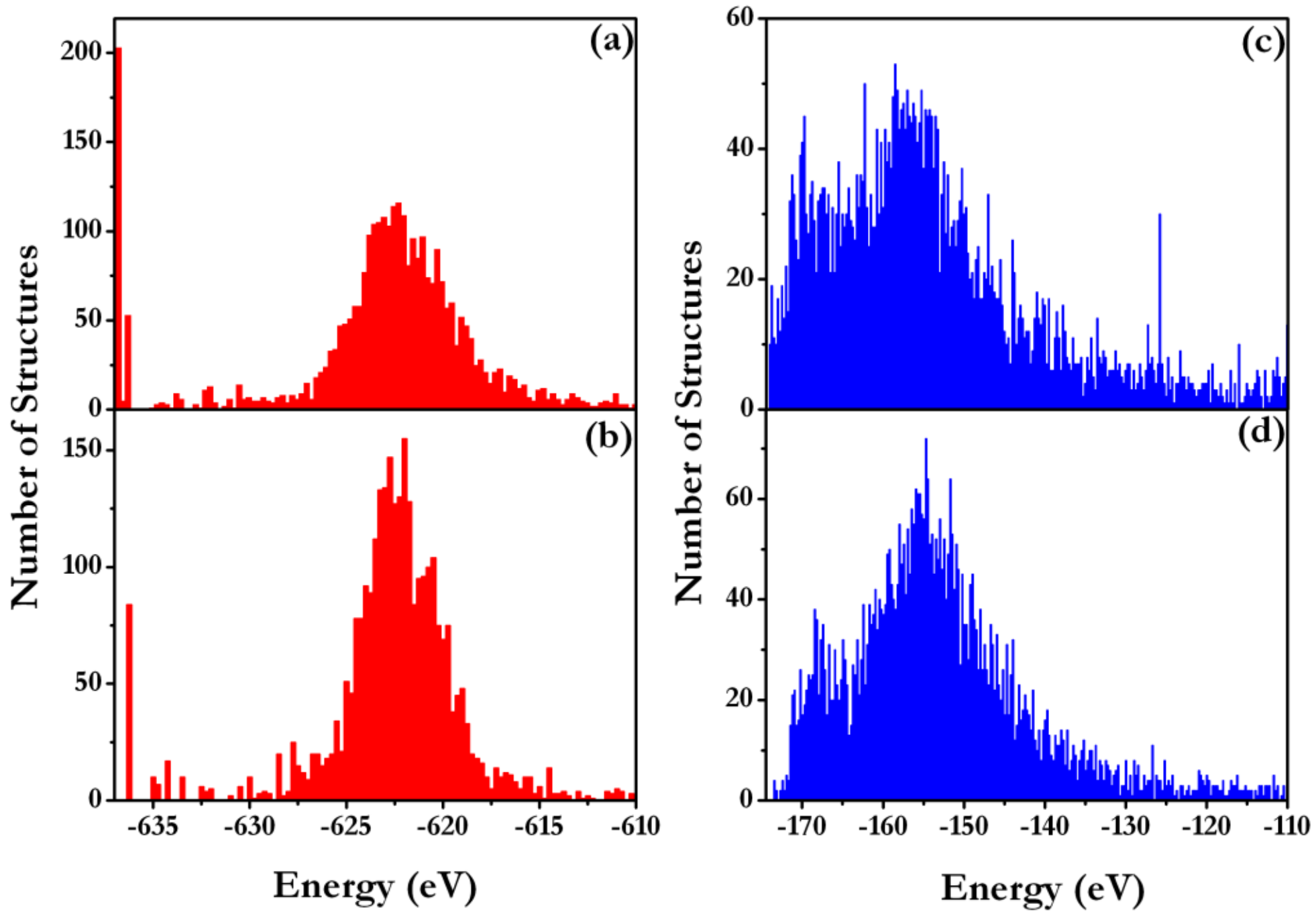

**Fig. 1**

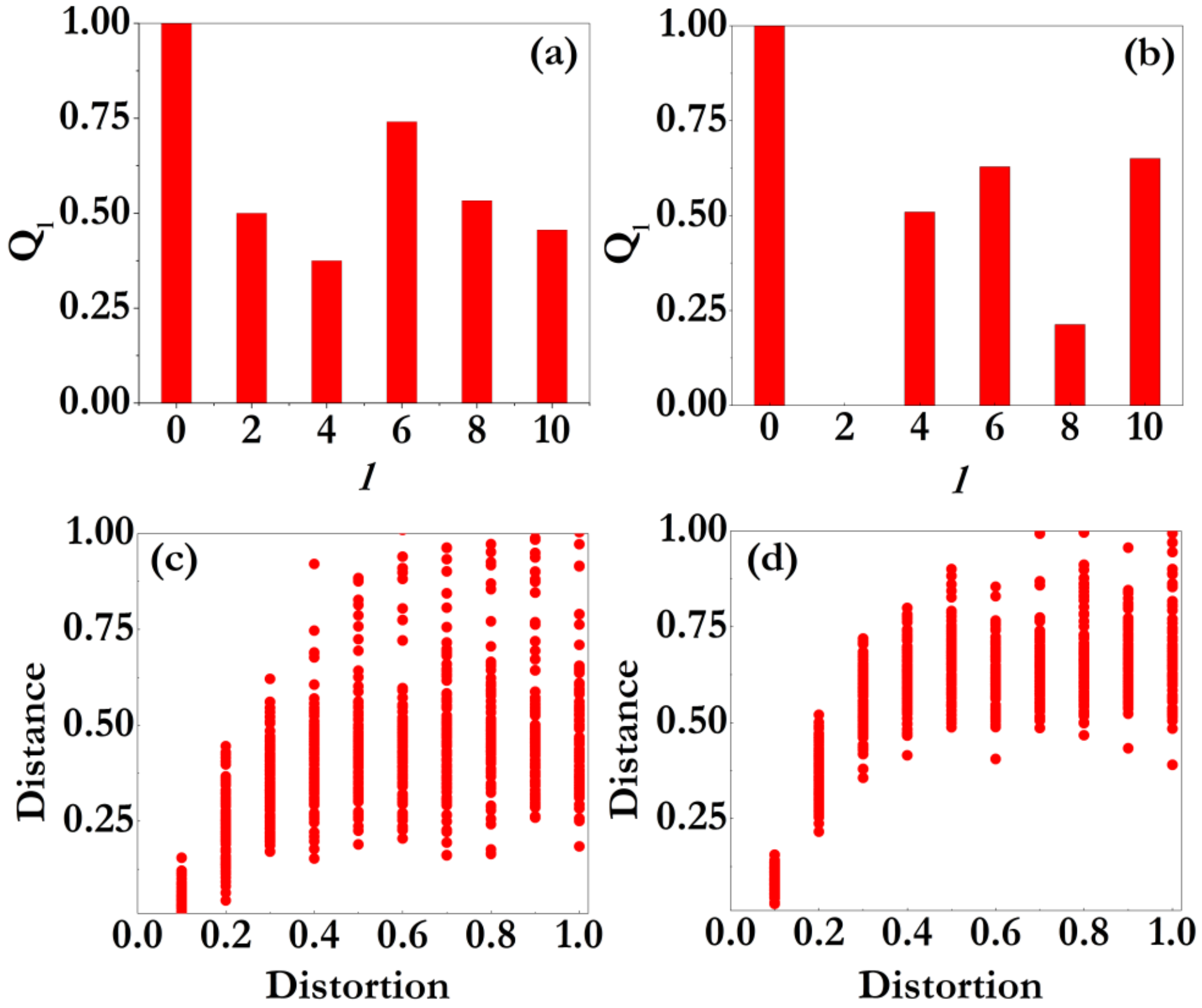

**Fig. 2**

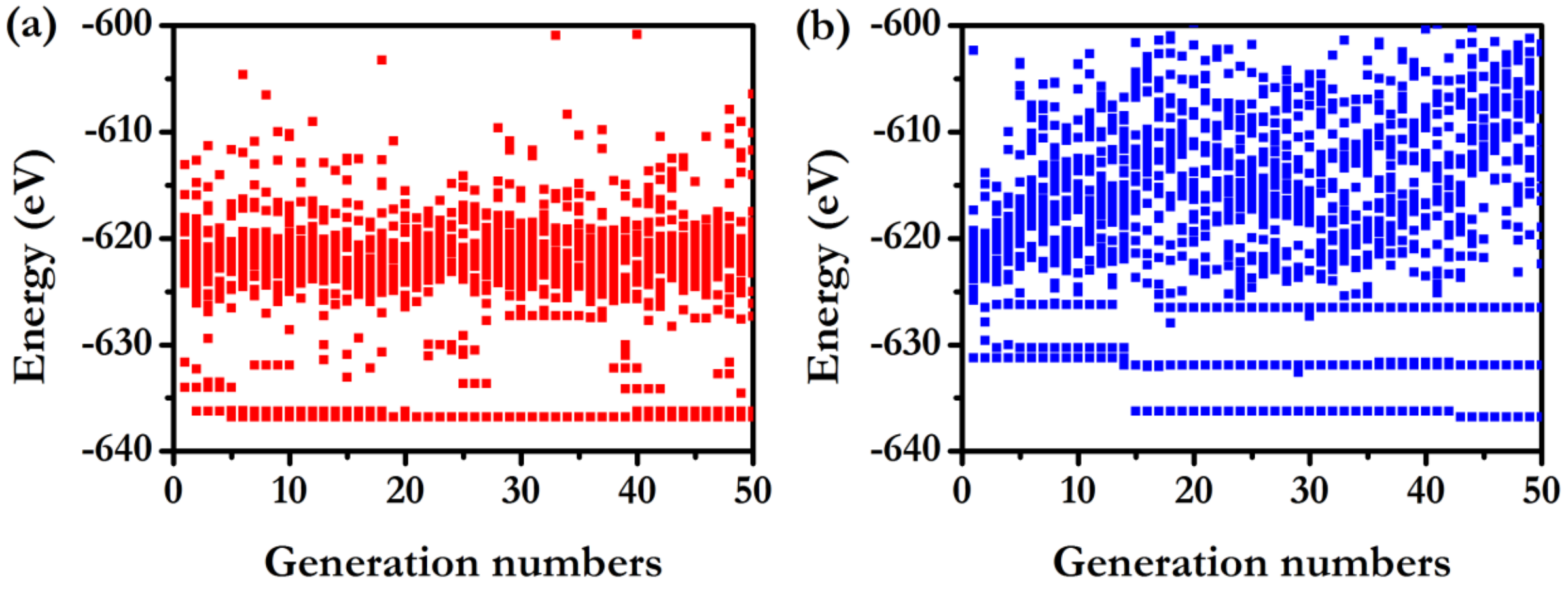

# Fig. 3

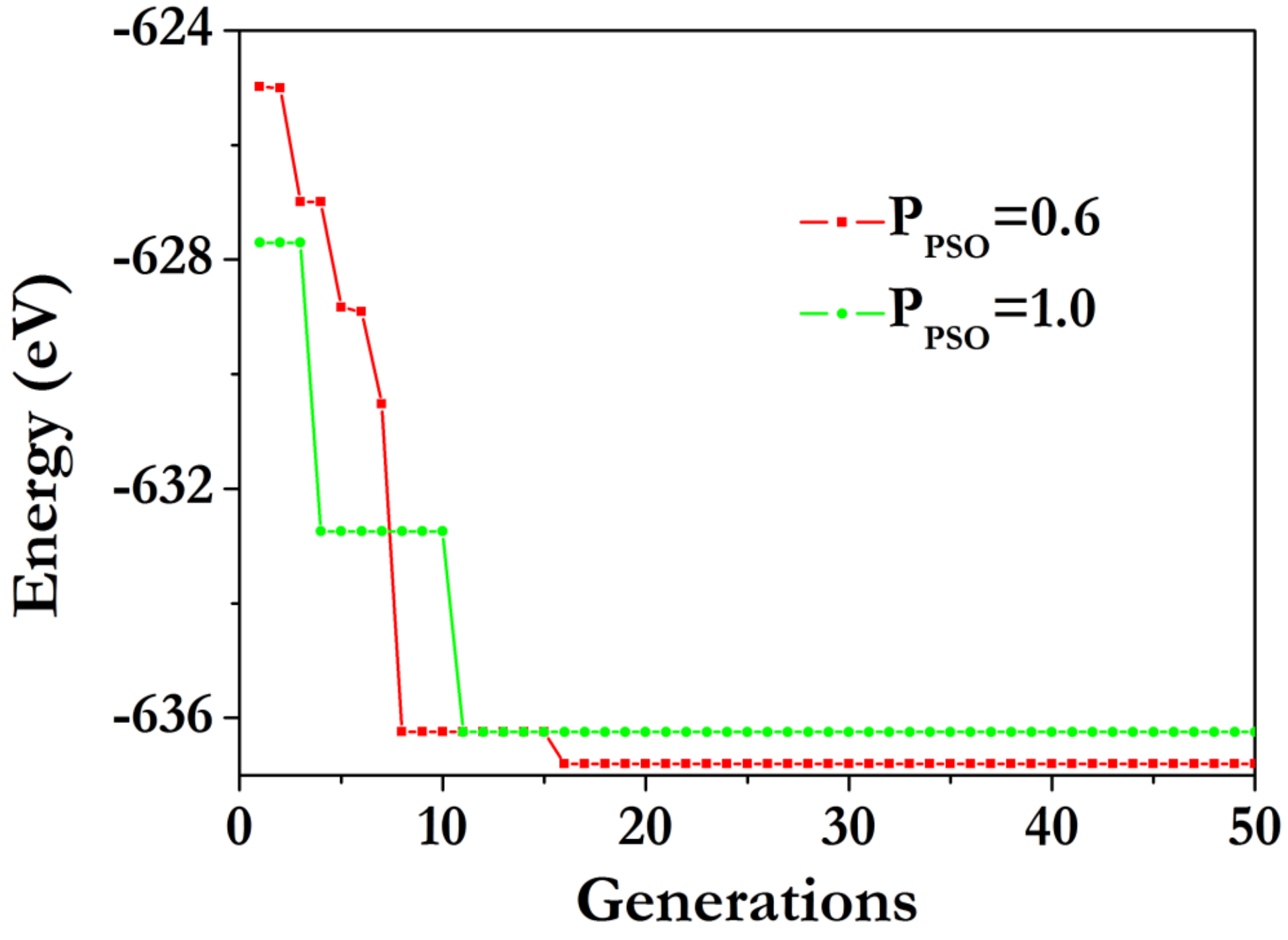

**Fig. 4**